\documentclass[twocolumn,showpacs,preprintnumbers,amsmath,amssymb]{revtex4}

\usepackage{graphicx}
\usepackage{dcolumn}
\usepackage{bm}
\usepackage{epsfig}

\def\bx{{\bf{x}}}
\def\bF{{\bf{F}}}

\newenvironment{Notes}
{\begin{quote}\small\tt Note to Daun: \ }
{\end{quote}}

\newcommand{\beq}{\begin{equation}}
\newcommand{\eeq}{\end{equation}}
\newcommand{\bno}{\begin{Notes}}
\newcommand{\eno}{\end{Notes}\noindent}

\begin{document}

\title{Reconstructing Equilibrium Entropy and Enthalpy Profiles from Non-equilibrium Pulling}

\author{Daun Jeong}
\author{Ioan Andricioaei}
 \email{andricio@uci.edu}
\affiliation{Department of Chemistry, University of California, Irvine, CA 92697}

\date{\today}

\begin{abstract}
The Jarzynski identity can be applied to instances when a microscopic system is pulled repeatedly but quickly along some coordinate, allowing the calculation of an equilibrium free energy profile along the pulling coordinate from a set of independent non-equilibrium trajectories. Using the formalism of Wiener stochastic path integrals in which we assign temperature-dependent weights to Langevin trajectories, we  derive exact formulae for the temperature derivatives of the free energy profile. This leads naturally to analytical expressions for decomposing a free energy profile into equilibrium entropy  and internal energy profiles from non-equilibrium pulling. This decomposition can be done from trajectories evolved at a unique temperature without repeating the measurement as done in finite-difference decompositions.  Three distinct analytical expressions for the entropy-energy decomposition are derived: using a time-dependent generalization of the weighted histogram analysis method, a quasi harmonic spring limit, and a Feynman-Kac formula. The three novel formulae of reconstructing the pair of entropy-energy profiles are exemplified by Langevin simulations of a two-dimensional model system prototypical for force-induced biomolecular conformational changes. Connections to single-molecule experimental means to probe the functionals needed in the decomposition are suggested.
\end{abstract}

\pacs{Valid PACS appear here}

\maketitle

\section{Introduction}
There exists for biomolecular systems (such as proteins or nucleic acids) substantial interest in calculating the equilibrium free energy difference $\Delta F$ between two states at equilibrium from repeated measurements of the non-equilibrium work $W$ done on the system along irreversible paths connecting the two states. One particular avenue, part of a broader class of approaches based on fluctuation theorems for systems out of equilibrium \cite{EvansCM93, EvansS94, GallavottiC95, Crooks99}, uses Jarzynski's non-equilibrium work theorem, $\exp(-\beta \Delta F) = \langle \exp(-\beta W) \rangle$, where the angular brackets denote averaging over the irreversible paths that start in one of the equilibrium states and $\beta$ is the inverse temperature \cite{JarPRL}. Hummer and Szabo have adapted the Jarzynski identity to show how one can calculate not just the free energy \textit{differences} $\Delta F$ between two states, but also entire free energy profiles $F(x)$ along some progress coordinate $x$ that is being pulled under non-equilibrium. They have also showed that this can be done not only on the basis of non-equilibrium molecular dynamics (MD) simulations, but also using data from actual single molecule pulling experiments \cite{HummerSzabo}. Examples involve the application of mechanical forces by laser or magnetic tweezers or by atomic force microscopy (AFM) to drive rare transitions in single molecules, such as unfolding of proteins and nucleic acid molecules, or the dissociation of a ligand from its protein target.  This is important because it allows the ability to compute free energy profiles along the direction of the pull in the absence of the pulling force. Such profiles can, in turn, provide valuable information about the kinetics and mechanisms of biomolecular folding or other important large-scale conformational transitions.  Further insight into the mechanisms of various conformational transitions can be provided if, in addition to the free energy profile, one could also compute its thermodynamics components: entropy and enthalpy, i.e., more precisely in the Helmholtz representation, the entropy profile $S(x)$ and the internal energy profile $U(x)$ in $F(x) = U(x) - T S(x)$.  Temperature-pertubation analytical formulae to decompose free energy profiles into entropy and enthalpy exist for equilibrium umbrella sampling \cite{PostmaBH82,SmithH93, WallqvistB95,LuKW03}, but not for the non-equilibrium Hummer-Szabo approach. In previous work \cite{NummelaYA08}, we derived formulae for entropy-energy decompositions for \textit{changes} between two states in the context of Jarzynski's identity. The purpose of this paper is to extend that work and present equations in the framework of the Hummer and Szabo approach for the calculation of entire entropy and energy \textit{profiles} based on an analytical approach derived from thermodynamic identities involving the temperature derivative of the free energy.  These derivatives are taken by introducing temperature dependance into the formula derived by Hummer and Szabo through the use of statistical re-weighting factors using stochastic path integrals techniques for Langevin dynamics.

\section{Theory}
\noindent The Jarzynski identity allows for the calculation of free energy profiles, as described by Hummer and Szabo \cite{HummerSzabo}, using the Feynman-Kac theorem, according to the formula
\begin{equation}
F(x)=-\frac{1}{\beta} \log \frac{
    \sum_t\frac{\langle \exp(-\beta W)\delta (x-x_t)\rangle}
                {\langle \exp(-\beta W)\rangle}
                            }
                           {
    \sum_t\frac{\exp(-\beta V_p(x,t))}
                {\langle \exp(-\beta W)\rangle}
                            }\rm,
\label{eqn:Fhs}
\end{equation}
where $x$ denotes the pulling molecular coordinate, $W=W[{\bf x}(t)]$ is the external work functional and $V_P= \frac{k}{2}(x-vt)^2$ is the pulling potential from the AFM cantilever moving with velocity $v$. 
This formula can be used both in simulations and experiments. In the simulation, multiple trajectories are run and the work values $W$ are computed; in the experiment, the work is integrated along the measured force-extension curves, which serve as one-dimensional trajectories.

The fundamental origin of the ability to decompose the free energy profile into entropy and energy profiles stems from the temperature dependence of the free energy profile $F(x)$: energy and entropy profiles can be calculated according to the basic thermodynamic relationships
\begin{eqnarray}
U(x)&=&\frac{\partial(\beta F(x))}{\partial \beta} \label{eqn:dU} \\
T S(x) &=& \beta \left(\frac{\partial F(x)}{\partial \beta}\right)\rm. \label{eqn:TdS}
\end{eqnarray}
Because $U(x)$ and $S(x)$ are $\beta$-derivatives of $F(x)$, if one can reconstruct $F(x)$ by recording trajectories at two nearby temperatures, one may estimate, in principle, the derivatives in Eqs. (\ref{eqn:dU}) and (\ref{eqn:TdS}) by a finite difference approximation (if one can detect the small signal needed for the finite differences from noise of the data) . However, it turns out that the derivatives can be estimated in closed form from trajectories recorded at a single temperature $\beta$. This is the gist of the present contribution: we show how, given a sufficiently large sampling of trajectories, and assuming Langevin dynamics at $\beta$, it is possible to calculate the free energy profile at any other value $\beta'$ in principle without approximation.  This can be done using Eq. (\ref{eqn:Fhs}), in which we multiply, {\`a} la umbrella sampling \cite{TorrieV77}, each functional value accumulated in the average estimator $\langle \cdot \cdot \cdot \rangle$ by a reweighting statistical factor, a functional $\Phi_{\beta'}[{\bf x}(t)]$ of the trajectory ${\bf x}(t)$.  This statistical factor will correspond to the probability of sampling a trajectory ${\bf x}(t)$ at the desired value of $\beta'$, divided by the probability of sampling the same trajectory at $\beta$.  In the case of Langevin dynamics, the trajectory probabilities can be expressed in terms of exponentials $\exp(-\beta \mathcal A_{\rm OM}[{\bf x}(t)] )$ of a stochastic action functional, $\mathcal A_{\rm OM}[{\bf x}(t)]$, the Onsager-Machlup action \cite{OnsagerM53,MachlupO53}.  An additional multiplicative factor of $\exp(-\beta V({\bf x}_0))$ accounts for the sampling measure of the initial equilibrium conditions.  Although the Onsager-Machlup action can be derived for a general form of the Langevin equation (even memory-dependent), let us assume, for simplicity an overdamped version. This is already a good approximation for molecular dynamics beyond the picosecond scale of biomolecular motion \cite{Jeremy}, which is well faster than the dynamics of relevance for important conformational changes. In the case of overdamped Langevin dynamics (with noise $\vec{ \xi}$ and friction $\gamma$ obeying fluctuation-dissipation),
\beq
  m \gamma \dot{\bf x} = -\nabla V({\bf x}) -k(x-vt) + \vec{ \xi}
\eeq
in a time-dependent force field ${\bf F}({\bf x},t) = -\nabla V({\bf x}) - k(x-vt)$ including both the systematic (molecular) force derived from the non-perturbed potential $V({\bf x})$ and the pulling harmonic force describing a virtual AFM moving cantilever with velocity $v$ along the pulling coordinate $x$, the Onsager Machlup action is the line integral $\mathcal A_{\rm OM} = (4m\gamma)^{-1} \int_0^{t} (m \gamma \dot{\bf x} -{\bf F}({\bf x},t))^2 dt $. The desired ratio of relative probabilities (i.e., the reweighting factor) then becomes
\begin{equation}
\Phi(\beta')=\exp((\beta-\beta')\mathcal A[{\bf x}(t)]),
\label{eq:Phi}
\end{equation}
where, in discrete form, the functional $\mathcal A=
\sum_{i=1}^n{
   (\frac{m \gamma}{4 \Delta t} \Delta \bx_i^2
    -\frac{1}{2} \Delta \bx_i \cdot \bF_i
    +\frac{\Delta t}{4 m \gamma} \bF_i^2)}+V(\bx_0)$.
Each thermodynamic average must then be divided by the average of the statistical factors.  This produces the formula
\begin{equation}
F(x,\beta')=-\frac{1}{\beta'}\log\frac{
    \sum_t\frac{\langle\exp(-\beta' W)\delta(x-x_t)\Phi(\beta')\rangle_\beta}
               {\langle\exp(-\beta' W)\Phi(\beta')\rangle_\beta}
                                            }
                                           {
    \sum_t\frac{\exp(-\beta' V_p(x,t))\langle\Phi(\beta')\rangle_\beta}
               {\langle \exp(-\beta' W)\Phi(\beta')\rangle_\beta}
                                            }\rm,
\label{eqn:Fofbeta}
\end{equation}
where the subscript $\beta$ on the angled brackets indicates that the averages are taken over a simulation or experiment run at $\beta$.

The $\beta'$ derivatives in Eqs. (\ref{eqn:dU}) and (\ref{eqn:TdS}) can then be taken directly.  Letting $f=\sum_t\frac{\langle\exp(-\beta' W)\delta(x-x_t)\Phi(\beta')\rangle_\beta}{\langle\exp(-\beta' W)\Phi(\beta')\rangle_\beta}$ and $g=\sum_t\frac{\exp(-\beta' V_p(x,t))\langle\Phi(\beta')\rangle_\beta}{\langle \exp(-\beta' W)\Phi(\beta')\rangle_\beta}$, this leads to 
\begin{equation}
U(x,\beta')=\frac{\partial(\beta' F(x,\beta'))}{\partial \beta'}
           =-\frac{\partial}{\partial\beta'}\log\frac{f}{g}=\frac{1}{g}\frac{\partial g}{\partial \beta'}-\frac{1}{f}\frac{\partial f}{\partial \beta'}
\end{equation}
With the notations
\begin{equation}
a(\beta')=\langle\exp(-\beta' W)\delta(x-x_t)\Phi(\beta')\rangle_\beta\rm,
\end{equation}
\begin{equation}
b(\beta')=\exp(-\beta' V_p(x,t))\langle\Phi(\beta')\rangle_\beta\rm,
\end{equation}
and 
\begin{equation}
c(\beta')=\langle\exp(-\beta' W)\Phi(\beta')\rangle_\beta,
\end{equation}
the derivatives evaluate to
\begin{equation}
U(x,\beta')=
  \frac{
    \sum_t
      \frac{c\frac{\partial b}{\partial\beta'}
            -b\frac{\partial c}{\partial\beta'}}
           {c^2}
        }
       {
    \sum_t b/c
        }
 -\frac{
    \sum_t
      \frac{c\frac{\partial a}{\partial\beta'}
            -a\frac{\partial c}{\partial\beta'}}
           {c^2}
        }
       {
    \sum_t a/c
        }
\label{eqn:Uofbeta-prime}
\end{equation}
By taking the derivatives of $a$, $b$ and $c$ in Eq. (\ref{eqn:Uofbeta-prime}) with respect to $\beta'$ explicitly, the expression can then be evaluated at $\beta'=\beta$ providing $U(x,\beta)$ and $TS(x)$ in terms of non-equilibrium trajectory averages taken at $\beta$:
\begin{eqnarray}
\label{eq:U}
U_{\rm HS}(x,\beta)&=&\frac{\sum_t
				\frac{\exp(-\beta V_p(x,t))}{\langle \exp(-\beta W)\rangle}
				[ -V_p-\langle \mathcal{A}\rangle+\langle\langle W+\mathcal{A}\rangle\rangle]}
			{\sum_t\frac{\exp(-\beta V_p(x,t))}{\langle \exp(-\beta W)\rangle}}\\ \nonumber
		&+&\frac{\sum_t \langle\langle\delta(x-x_t)(W+\mathcal{A})\rangle\rangle-
				\langle\langle\delta(x-x_t)\rangle\rangle\langle\langle W+\mathcal{A}\rangle\rangle}
			{\sum_t \langle\langle\delta(x-x_t)\rangle\rangle},\\
	TS_{\rm HS}(x,\beta)	&=&  U_{\rm HS}(x,\beta) - F_{\rm HS}(x,\beta) 	
\end{eqnarray}
where $\langle\langle \cdots \rangle\rangle$ is evaluated as the work-weighted average such that $\langle (\cdots) \exp(-\beta W)\rangle/\langle \exp(-\beta W)\rangle$, 
and the entropy profile is expressed as the difference between the analytical expressions for internal energy and free energy profiles above. Although somewhat more tediously, the same expressions can be obtained by explicitly showing the $\beta$-dependence of all the averages $\langle \cdot \cdot \cdot \rangle$ in Eq. (\ref{eqn:Fhs}) when the averages are written as path integrals over all possible trajectories with their ($\beta$-dependent) measure written explicitly, followed by taking all partial $\beta$-derivatives needed for Eqs. (\ref{eqn:dU})-(\ref{eqn:TdS}).

Instead of using the weighted-histogram analysis method in Eq.~(\ref{eqn:Fhs}), Hummer and Szabo recently developed a quasi-harmonic approximation formula for the free energy profile as a function of the molecular pulling coordinate \cite{HummerS10}
\begin{eqnarray}
F_{\rm QH}\left(x=vt-\frac{\langle\langle F_p\rangle\rangle}{k}\right)\approx -\frac{1}{\beta}\ln \langle \exp(-\beta W)\rangle \\ \nonumber
-\frac{\langle\langle F_p\rangle\rangle ^2}{2k}
+\frac{1}{2\beta}\ln\left( \beta(\langle\langle F_p^2\rangle\rangle-\langle\langle F_p\rangle\rangle^2)/k\right),
\end{eqnarray} 
where $F_p$ denotes the pulling force $-k(x-vt)$. 
Following the same procedure to derive Eq.~(\ref{eq:U}), we obtain the corresponding quasi-harmonic approximation for the energy $U_{\rm QH}(x)$ and entropy $S_{\rm QH}(x)$ profiles, as below: 
\begin{eqnarray}
\label{eq:U_QH}
&&U_{\rm QH}\left(x=vt-\frac{\langle\langle F_p\rangle\rangle}{k},\beta\right)
\approx 
		\langle\langle W+\mathcal{A}\rangle\rangle-\langle \mathcal{A}\rangle \\ \nonumber
		&-&\frac{\langle\langle F_p\rangle\rangle^2}{2k}
		-\frac{\beta}{k}\langle\langle F\rangle\rangle [\langle\langle F_p(W+\mathcal{A})\rangle\rangle-\langle\langle F_p\rangle\rangle\langle\langle W+\mathcal{A}\rangle\rangle] \\ \nonumber
		&+&\frac{1}{2\beta}
		+\frac{\langle\langle F_p^2(W+\mathcal{A})\rangle\rangle-\langle\langle F_p^2\rangle\rangle\langle\langle W+\mathcal{A}\rangle\rangle}
		{2[\langle\langle F_p^2\rangle\rangle-\langle\langle F_p\rangle\rangle^2]}\\ \nonumber
	       &-&\frac{\langle\langle F_p\rangle\rangle[\langle\langle F_p(W+\mathcal{A})\rangle\rangle-\langle\langle F_p\rangle\rangle\langle\langle W+\mathcal{A}\rangle\rangle]}
			{[\langle\langle F_p^2\rangle\rangle-\langle\langle F_p\rangle\rangle^2]}\\
&&TS_{\rm QH}\left(x=vt-\frac{\langle\langle F_p\rangle\rangle}{k},\beta\right) = 	U_{\rm QH} - F_{\rm QH}	
\end{eqnarray}

Expressions for the energy-entropy profiles along $x$ can also be derived  using the thermodynamic relations and the Feynman-Kac theorem in the same manner that $F(x)$ in our Equation~(\ref{eqn:Fhs}) was obtained in Ref.~\onlinecite{HummerSzabo},
\begin{equation}
U_{\rm FK}(x)=\frac{\langle H_t\exp(-\beta W)\delta(x-x_t)\rangle}{\langle \exp(-\beta W)\delta(x-x_t)\rangle} - \langle H\rangle,
\end{equation}
where $H_t=V({\bf x}_t)+m\dot{\bf x}_t^2/2$. This formula does not require the calculations of trajectory weights. Adapting the weighted histogram method, the energy profile up to an additive constant is computed as 
\begin{eqnarray}
U_{\rm FK}(x)&=&\frac{
    \sum_t\frac{\langle V({\bf x}_t)\exp(-\beta W)\delta (x-x_t)\rangle}
                {\langle \exp(-\beta W)\rangle}
                            }
                           {
    \sum_t\frac{\langle \exp(-\beta W)\delta (x-x_t)\rangle}
                {\langle \exp(-\beta W)\rangle}
                            }\\ \label{eq:U_FK}
TS_{\rm FK}(x) &=& U_{\rm FK}(x) - F(x).                            
\end{eqnarray}

\section{Numerical Demonstration}
\begin{figure}[h]
  \centerline{\epsfig{figure=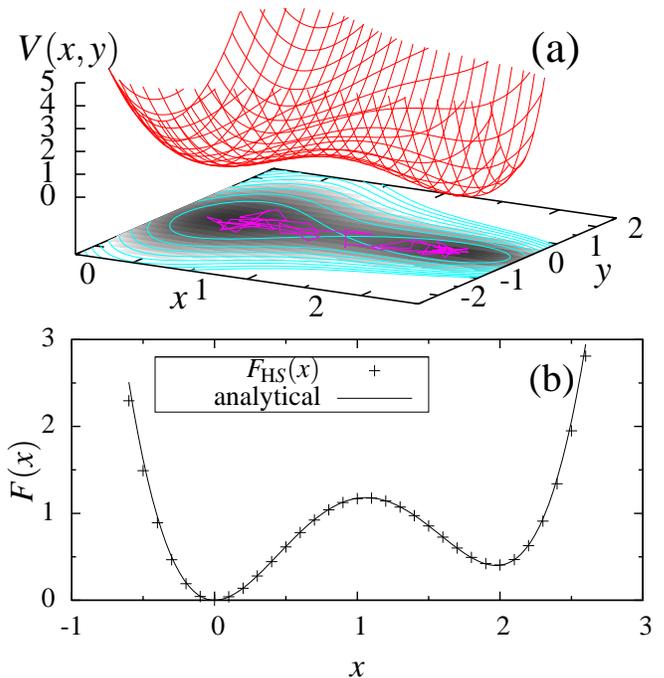,width=3.5in}}
  \caption{(a) Two-dimensional model potential prototypical of biomolecular conformational changes, with conformational transition occurring along $x$ across a barrier, and with a harmonic perpendicular degree of freedom $y$ with decreasing width (decreasing entropy) as $x$ increases. This could model an ``unfolded" state at $x=0$ and a ``folded" state at $x=2$. (b) Reconstructed (+) and analytical (solid line) free energy profiles.}
  \label{fig:F}
\end{figure}

A model potential to test the method is selected for which free energy, energy and entropy profiles can be calculated analytically:
\begin{equation}
V(x,y)=p(x)+q(x) y^2,
\end{equation}
with $p(x)=x^2(x-2)^2$ and $q(x)=(x^2+1)$, which is depicted in Fig.~\ref{fig:F}(a).  
The potential corresponds to a bistable potential in the $x$ direction, with minima at $x=0$ and $x=2$, joined by a potential energy barrier of height $1$ at $x=1$.  In the $y$ direction the potential is a harmonic oscillator with a force constant that varies with $x$, akin to a trough aligned along $x$ of increasing curvature and with a bump in the middle.  The minima at $(x=2,y=0)$ is entropically destabilized relative to the minima at $(x=0,y=0)$ because the latter well is wider along $y$. This 2-dimensional energy is perhaps the simplest model for a biomolecular conformational transition (such as a two-state protein folder) between two states separated by a barrier, and with the free energy in the ``important" direction $x$, (e.g., the folding coordinate) modulated by entropy.

The partition function for this model as a function of the $x$ coordinate is
\begin{equation}
Q(x)=2 m (\pi / \beta)^{3/2} \frac{e^{-\beta p(x)}}{\sqrt{q(x)}}.
\end{equation}
From the partition function free energy ($F(x)$), internal energy ($U(x)$), and entropy ($S(x)$) profiles can be calculated according to the relations $F=- \beta^{-1} \log Q$, $S=-(\frac{\partial F}{\partial T})$, and $U=F+T S$.  The corresponding analytic results for these profiles are
\begin{equation}
F(x)=p(x)-\frac{1}{\beta}\log\frac{2 m (\pi / \beta)^{3/2}}{\sqrt{q(x)}},
\end{equation}
\begin{equation}
T S(x)=\frac{3}{2\beta}+\frac{1}{\beta}\log\frac{2 m (\pi / \beta)^{3/2}}{\sqrt{q(x)}},
\end{equation}
and
\begin{equation}
U(x)=p(x)+\frac{3}{2 \beta}.
\end{equation}
To carry out a reconstruction of the energy profiles along the $x$ coordinate, a harmonic pulling potential of the form $V_p(x)=\frac{k}{2}(x-v t)^2$ is superimposed on the underlying potential, where $k$ is the spring constant, $v$ is the pulling velocity and $t$ is time.

Overdamped Langevin trajectories were simulated on the potential described above, with parameters $\gamma$ and $m$ chosen to be unity, and $\beta$ set to be $2$; a number of $10^6$ trajectories were integrated for $10^4$ time steps of size $.001$ in arbitrary units.  The spring constant, $k$, used for the pulling potential was $5$, and a pulling velocity of $.2$ was used so that the minima of the pulling potential would vary from $0$ to $2$ over the course of each trajectory. When evaluating the quasi-harmonic approximation, $k$ was set to $10$, which yielded smaller deviations near the barrier. 

\begin{figure}[h]
  \centerline{\epsfig{figure=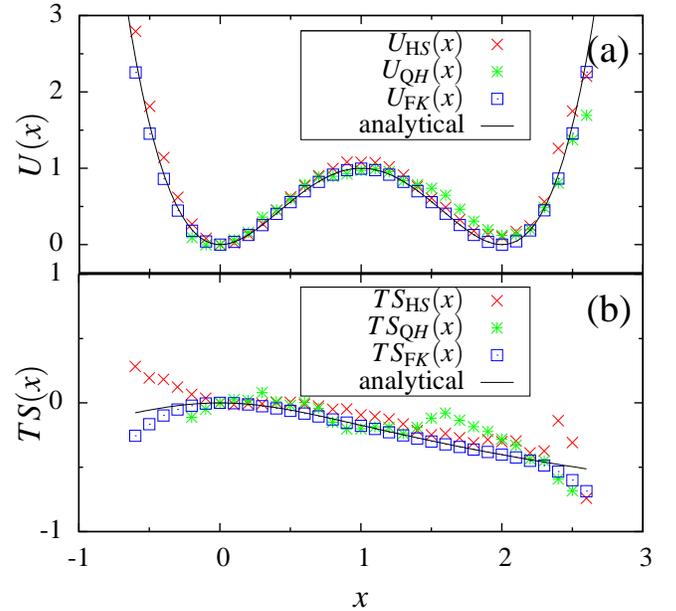,width=3.5in}}
  \caption{(a) Energy profiles reconstructed from Eq.~(\ref{eq:U}), (\ref{eq:U_QH}), and (\ref{eq:U_FK})   (symbols) and analytical result (solid line) (b) Reconstructed (symbols) and analytical (solid line) entropy profiles. }
  \label{fig:US}
\end{figure}

Free energy, energy and entropy profiles are all recovered with reasonable agreement to the analytic result.  The free energy profile is exact (Fig.~\ref{fig:F}(b)).  
Energy profiles calculated using three different expressions in Eq.~(\ref{eq:U}), (\ref{eq:U_QH}), and (\ref{eq:U_FK}) are shown in Fig.~\ref{fig:US}(a). $U_{\rm HS}(x)$ and $U_{\rm QH}(x)$ show some deviation at poorly sampled values of $x$, around the barrier and outside of the interval $0$ to $2$, while $U_{\rm FK}(x)$ agrees with the theoretical values derived analytically. Entropy profiles shown in Fig.~\ref{fig:US}(b) show similar behavior. $U_{\rm QH}$ and $S_{\rm QH}$ in particular have failed to converge for values of $x$ over the barrier.  All profiles have been arbitrarily aligned at $x=0$.

\begin{figure}[h]
  \centerline{\epsfig{figure=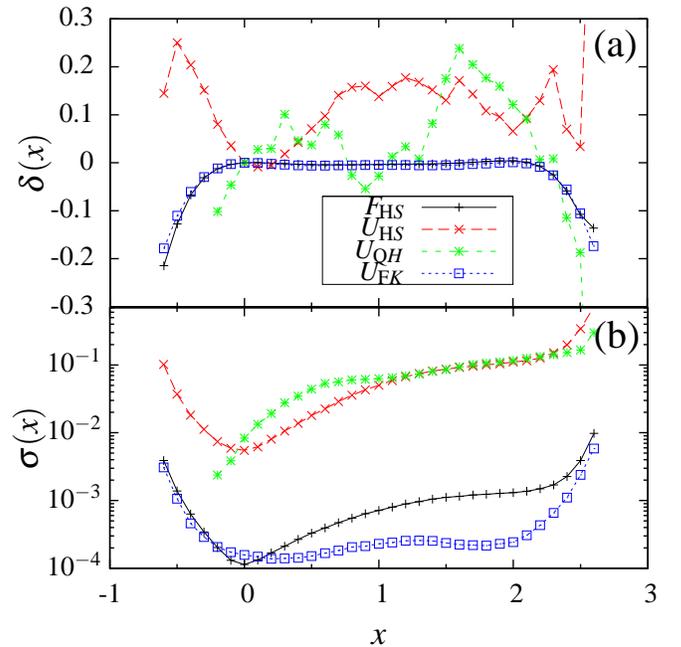,width=3.5in}}
  \caption{ (a) Biases $\delta(x)$ and (b) standard deviation $\sigma(x)$ for each methods used to calculate $F(x)$ and $U(x)$. }
  \label{fig:delta}
\end{figure}

The accuracy of the results shown in Fig.~\ref{fig:F} and \ref{fig:US}(a) is measured by calculating bias given as $\delta(x)=\hat{\overline{u}}(x)-u(x)$, where  $\hat{\overline{u}}(x)=1/N\sum_{n=1}^N u_n(x)$ and $u_n(x)$ is the value of $n$-th sample for $F_{\rm HS}(x)$, $U_{\rm HS}(x)$, $U_{\rm QH}(x)$, and $U_{FK}(x)$, computed from 1000 trajectories, while $u(x)$ is the exact values of the profiles derived analytically. $N$ is the number of independent samples and set to be 1000.  
On the other hand, the standard deviation $\sigma(x)=(1/N\sum_{n=1}^N [u_n(x)-\hat{\overline{u}}(x)]^2)^{1/2}$ reveals the relative precision of the values calculated with each methods. 
$\delta(x)$ and $\sigma(x)$ corresponding to four profiles are displayed in Fig.~\ref{fig:delta}. 

In cases of $F_{\rm HS}(x)$ and $U_{\rm FK}$, biases are negligible compared with those of $U_{\rm HS}(x)$ and $U_{\rm QH}(x)$, which were computed using the trajectory weights. Likewise, the standard deviations for the former are smaller by two orders of magnitude than those of the latter in general. 
The large values of $\sigma(x)$ for the latter arise from the large variance of trajectory weights, and the increasing behavior is because of less sampling of $x$ and the accumulated error in the calculations of trajectory weights.

\section{Conclusions}
We have presented formulae, analogous to the free energy profile formula of Hummer and Szabo, to calculate energy and entropy profiles.  The formulae can be applied rigorously to driven MD simulations of biomolecules to providing insight into driving forces for conformational changes. Although not more accurate than equilibrium decompositions of free energy on the basis of thermodynamic integration or thermodynamic perturbation \cite{NummelaYA08}, an important advantage of the decomposition of free energy from non-equilibrium simulations over equilibrium simulations is that they are trivially parallel, hence many quick simulations can be distributed on many independent CPUs and run at much shorter wall-clock time. However, while application to simulations is straightforward, application to single molecule pulling experiments is complicated by two restrictive aspects. 

The first aspect is general to reconstruction of both the energy-entropy profile decomposition and to the original free energy profile reconstruction and has to do with the fact that exact formulae are expressed in terms of infinitely many trajectory realizations. This is in contrast with experimental situations, for which only finite numbers of trajectory measurements are available. This convergence issue is particularly acute when the pulling coordinate is varied much more rapidly than the equilibration time of the coordinate. In these far out-of-equilibrium conditions, the averages such as those derived for internal energy, entropy or free energy can be dominated by realizations that are extremely rare, and more and more trajectories are likely to be needed. 

The second complicating aspect has to do with the ability to monitor ``perpendicular" degrees of freedom, i.e., other important degrees of freedom than the end-to-end extension that is being pulled. Measurement of the action $\mathcal A$ in Eq. (\ref{eq:Phi}) needed to reweight requires and estimate of the multidimensional force $\bF$ and displacement $\Delta \bx$. To work in resolving entropy profiles, more than one degree of freedom would need to be probed (because, for a one-dimensional Langevin propagation on the reconstructed free energy, the 1-d free energy profile is also the potential-of-mean-force profile (whose gradient is the effective force), hence the entropy profile is constant).  While the pulling experiments typically can report forces and displacement along a single direction (and have been used in the context of fluctuation theorems to reconstruct free energies \cite{LiphardtDSTBusta2002,CollinRJSTB05,HarrisSK07}), it is possible in principle  in single molecule experiments to assess the effect of different pulling geometry on the mechanical strength; this has been reported in two circumstances that we are aware of. 

Firstly, while most proteins studied to date involve pulling recombinant tandem arrays of homopolymers, which limits the extension geometry to that applied between the N- and C-termini, other extension geometries are possible. Examples are studies of a protein, E2lip3, specifically labeled with a gold reactive tag at specific sites \cite{BrockwellPZBOSPR03}; ubiquitin, which can form polymers between its C-terminus and the side chain of one of four lysine residues \cite{Carrion-VazquezLLMOF03}; and lysozyme, which has been linked by novel disulphide bonds \cite{YangCBVBHMDB00}. 

Secondly, instead of (or in addition to) pulling from different ends of the same proteins as done in the experiments above,  another experimental procedure of use would be to probe N- to C- terminus separation by pulling \textit{permutants} of the protein, i.e., mutants that have the same sequence but different positions of the N and C termini; such a strategy is possible and was used to study the effect of protein structure on mitochondrial import \cite{WilcoxCBM05}.

Such type of force-spectroscopy investigations open the possibility to build approximate models for what exactly fluctuates in the perpendicular manifold (i.e., which structural degrees of freedom collectively contribute to entropy as described by a one-dimensional perpendicular harmonic variable). In principle, this harmonic motion could then be probed spectroscopically during the single-molecule pulling experiment, and this strategy can lend the decomposition formulae presented herein to useful applications. Possible experimental scenarios might involve suitably placed two- or three-color FRET dyes \cite{Xie02,RasnikMH05}  or polarization-sensitive spectroscopy with a rigidly attached dye \cite{Fourkas01} that can be used to gauge time-dependent perpendicular fluctuations  concomitant to pulling.




While the formulation here was based on the original Hummer and Szabo approach, it may be of interest to explore the decomposition of entropy and internal energy in the framework of other approaches to free energy reconstruction, such as the extensions to the fluctuation theorem to include reaction coordinates \cite{ParamoreAV07} or the differential fluctuation theorem \cite{ MaragakisSK08}.


\section{Acknowledgments}  IA acknowledges funds from an NSF CAREER award (CHE-0548047).

\clearpage
\bibliographystyle{biophysj}
\bibliography{./ges,../../all}
\end{document}